\documentclass [a4paper,twoside,11pt]  {article}
\usepackage {a4wide,amsmath,amssymb,latexsym}
\usepackage {delarray,graphics,graphpap,alltt,graphicx}
\usepackage{setspace}  % four double spacing
\usepackage {a4wide,amsmath,amssymb,latexsym,graphicx}
\usepackage{rotating}
\newcommand{\mca}{\mathcal}

\newcommand{\N}{\mathbb{N}}

\newcommand{\R}{\ensuremath{\mathbb{R}}}

\newcommand{\Z}{\ensuremath{\mathbb{Z}}}

\def\squareforqed{\hbox{\rlap{$\sqcap$}$\sqcup$}}
\def\qed{\ifmmode\squareforqed\else{\unskip\nobreak\hfil
\penalty50\hskip1em\null\nobreak\hfil\squareforqed
\parfillskip=0pt\finalhyphendemerits=0\endgraf}\fi}

\newenvironment{proof}{  \noindent\textit{Proof}.}{} % on peut remplacer {} par {\qed} pour forcer un carre
\newtheorem{definition}{Definition}[section]
\newtheorem{theorem}{Theorem}[section]
\newtheorem{lemma}{Lemma}[section]
\newtheorem{corollary}{Corollary}[section]

\newcounter{myenumi}
  {\begin{list}{\textup{(\arabic{myenumi}) }}{\usecounter{myenumi}%
    \setlength{\labelsep}{0pt}\setlength{\leftmargin}{0pt}%
    \setlength{\labelwidth}{0pt}%
    \setlength{\listparindent}{0pt}}}
  {\end{list}}

\newcounter{myenumii}
  {\begin{list}{\textup{(\arabic{myenumi}.\arabic{myenumii}) }} {\usecounter{myenumii}%
    \setlength{\labelsep}{0pt}\setlength{\leftmargin}{0pt}%
    \setlength{\labelwidth}{0pt}%
    \setlength{\listparindent}{0pt}}}
  {\end{list}}

\newcommand{\lun}{L^1_{\mathrm{rec}}}
\newcommand{\lpm}{L^p}
\newcommand{\lp}{L^p_{\mathrm{rec}}}
\newcommand{\noun}[1]{\| #1 \|_1}
\newcommand{\nop}[1]{\| #1 \|_p}
\newcommand{\comment}[1]{}
\title{On the Convergence of Fourier Series of Computable Lebesgue Integrable Functions}
\author{Philippe Moser
\footnote{Dept. de Inform\'atica e Ingenier\'{\i}a de Sistemas,
Edificio Ada Byron, Mar\'{\i}a de Luna 1, 50018 Zaragoza, Spain. Email: mosersan (at) gmail.com. Research supported in part by Spanish Government MEC Program Juan de la Cierva.
}}
\date{}

\begin{document}
	\maketitle
%\doublespacing
%\begin{spacing}{3} % double space for easy correction
%%%%%%%%%%%%%%% abstract %%%%%%%%%%%%%%%%%%%%%%%%%%%%%%%%%%%%%%%

	\begin{abstract} 
		This paper studies how well computable functions
		can be approximated by their Fourier series.
		To this end, we equip the space of $L^p$-computable functions (computable Lebesgue integrable functions) 
		with a size notion, by introducing $L^p$-computable Baire categories.
		We show  that $L^p$-computable Baire categories satisfy the following three basic properties.
		Singleton sets $\{f\}$ (where $f$ is $L^p$-computable) are meager, suitable infinite unions of meager sets are meager,
		and the whole space of $L^p$-computable functions is not meager.
		We give an alternative characterization of meager sets via Banach Mazur games.
		We study the convergence of Fourier series for $L^p$-computable functions and show that whereas for every $p>1$, the Fourier series
		of every $L^p$-computable function $f$ converges to $f$ in the $L^p$ norm, the set of $L^1$-computable functions  whose Fourier series
		does not diverge almost everywhere is meager.
	\end{abstract}

\section{Introduction}

	Fourier series are trigonometric 
	polynomials that are useful for approximating arbitrary periodic functions. Areas of applications include electrical engineering,
	acoustics, optics, signal and image processing, and data compression. The goal of this paper is to study how well computable functions
	can be approximated by their Fourier series. Our main result shows that almost all computable Lebesgue integrable functions cannot 
	be approximated by their Fourier series. 

	Our work is based in the  setting of computable Lebesgue integrable functions 
	(see \cite{b.pour-el.computable.analysis,b:weihrauch.computable.analysis.introduction}), a natural extension of the standard 
	bit-computable (BC) model (see \cite{b.pour-el.computable.analysis}), where a function $f$ is said computable if a 
	TM given a good approximation for $x$, can compute 
	a good approximation for $f(x)$. One of the limitation of the BC model, is the fact that every bit-computable function is continuous. 
	Therefore even simple step functions
	-- a tool extensively used in functional analysis -- are not computable in the BC model.
	A natural extension of the BC model known as $L^p$-computability, consists of  a computable version of $L^p$  
	(see \cite{b.pour-el.computable.analysis,b:weihrauch.computable.analysis.introduction}), 
	the space of $p$-power Lebesgue integrable functions (where $L^p$ is the class of functions $f$ such that 
	$|f|^p$ is Lebesgue integrable). In functional analysis, the  spaces $L^p$ form an important class 
	of examples of Banach spaces.
	The notion of $L^p$-computable functions can be seen as an extension of the BC-model, which informally
	corresponds to the computable version
	of the class of continuous functions.

	In order to be able to prove quantitative results (i.e. of the form: \emph{almost every} function $f$ satisfies $\ldots$),
	we equip the space of $L^p$ computable functions with a size notion, by introducing $L^p$-computable Baire categories.
	Classically (see \cite{b:oxtoby}), Baire categories are a topological size notion that allow to characterize the size 
	of subsets of $L^p$, and that satisfy the following three basic properties.
	Singleton sets $\{f\}$ ($f\in L^p$) are meager (i.e. small), countable unions of meager sets are meager,
	and the whole space $L^p$ is not meager. Baire categories can be used to show quantitative results 
	instead of existential ones, i.e.
	show that most functions have some property $P$ instead of proving the mere existence of one such  function. 
	Unfortunately, classical Baire categories cannot be used directly as  a size notion on the set of computable functions,
	because this set is countable hence meager. What is needed is a computable version of Baire categories. We introduce such 
	a notion on the space of $L^p$-computable functions, and show that our notion satisfies the three basic properties.
	Classically, meager sets can be characterized via Banach Mazur games (infinite two players games) (see \cite{b:oxtoby}).
	We show that a similar characterization of meagerness holds for our $L^p$-computable Baire categories.

	Our work extends previous notions of Baire categories introduced in the more restricted setting of bit-computability
	\cite{DBLP:conf/focs/Mehlhorn73,b:lutz-baire-category-for-real-functions}.

	We then investigate the convergence of Fourier series for $L^p$-computable functions. We use  our Baire category notion to understand 
	how well can $L^1$-computable functions be approximated by their Fourier series. 
	It is well know that for $p>1$,  the Fourier series of any function
	in $L^p$ converges to $f$ in the $L^p$-norm, and this is true for $L^p$-computable functions also.
	But for $p=1$, things change dramatically: In the early 20's, 
	Kolmogorov \cite{b:kolmo.counter-example.in.L1} constructed
	a function $f\in L^1$ whose Fourier series diverges almost everywhere. Unfortunately, Kolmogorov's result
	gives no information in the setting of $L^1$-computable functions. 

	We show that the analogue is true in the computable case, 
	and that it is a \emph{typical} property of $L^1$-computable functions, i.e. a majority of $L^1$-computable
	functions are very ``complicated'' functions, in the sense that most of them cannot be approximated by their Fourier series.
	More precisely we show that 
	the set of $L^1$-computable functions whose Fourier series does not diverge a.e. is $L^p$-meager, i.e. negligible.
	As   corollaries, our result implies the existence of an $L^1$-computable function  whose Fourier series diverges almost everywhere,
	as well as Kolmogorov's result \cite{b:kolmo.counter-example.in.L1}.

\section{Preliminaries}
	We take as known the elements of Lebesgue's theory of measure and integration, see for example 
	\cite{b:zygmund.trigo.series.vol.1,b:hardy.Fourier.series}.
	If not mentioned, all functions we consider are over $[0,2\pi]$.
	$\R^{[0,2\pi]}$ denotes $\{f| \, f:[0,2\pi]\rightarrow \R\}$.
	$L$ is the set of functions $f(x)$ integrable in the sense of Lebesgue over $[0,2\pi]$.
	A measure zero set is called a \emph{null set}: null sets are irrelevant in the theory of integration.
	We say $f=g$ \emph{almost everywhere} (a.e.) if $f$ and $g$ differ only on a null set.
	For $p\geq 1$, we say $f$ is in $L^p$ if $f$ is measurable and $|f|^p$ is in $L$. $L^1$ is $L$.
	The \emph{$p$-norm} is 
	\[
		\nop{f} = \left( \int_{0}^{2\pi} |f|^p \, dx \right)^{1/p}
	\] 
	$L^p$ is a metric space, with the distance given by
	\[
		d_p(f,g)=\nop{f-g}.
	\]
	We will not distinguish two functions that are equal a.e., and use the notation $f$ for representing the class of functions
	equal a.e. to $f$.

\section{The Class of $L^p$-computable Functions}
	
	A function $f$ is said $L^p$-computable  if a Turing machine can compute a step function approximating $f$ in the $p$-norm. More precisely,
	\begin{definition} [see \cite{b.pour-el.computable.analysis,b:weihrauch.computable.analysis.introduction}]
		Let $p\in\Z^+$.
		A (class of) function $f:[0,2\pi]\rightarrow \R$  is in $\lp$ (also said $L^p$-computable) if there exists a TM  that on input any 
		$m\in\N$, outputs a set of rationals $\{(a_i,b_i)|\, 1\leq i \leq t_m\}$ such that the step 
		function $\psi|_{[a_i,a_{i+1})} = b_i \quad (1\leq i\leq t_m -1)$ satisfies $\|\psi-f\|_p < 2^{-m}$.
 	\end{definition}

	The above step function $\psi$ is called a rational step function with endpoints $\{(a_i,b_i)|\, 1\leq i\leq t_m\}$.
	For families of functions the definition is similar: a family of functions $\{f_n\}_{n\geq 0}$ is in $\lp$ (or is uniformly $L^p$-computable) 
	if there is a  TM
	that on input any two integers $m,n$, outputs integers corresponding to a step function
	$\psi$ such that $\|\psi-f_{n}\|_p < 2^{-m}$. 

	More generally we call a family of objects $\{O_{n}\}_{n\geq 0}$ 
	(rational numbers, functions, etc.) to be uniformly computable if there is a single TM that given $n$ computes $O_{n}$.

\comment{\subsection{Comparison with other Notions}
	
	It is clear that bit-computability implies $L^p$-computability. The reverse is not true because rational step functions
	are not bit-computable. On the other hand $L^p$-computability and graph-computability  (a function is graph computable
	\cite{b:Braverman-Complexity-Real-Functions} if 
	its graph $\Gamma (f) = \{(x,f(x))|\, x\in [0,2\pi]\}$ is a computable set)
	are incomparable as the following result shows.

	\begin{theorem} 
	Graph-computability and $L^p$-computability are incomparable ($p\in\Z^+$).
	\end{theorem}
	\begin{proof}
		Let $p\in\Z^+$. Take the function 
	\[
		f(x)=
		\begin{cases}
			0 &\text{ if $x\in [0,2\pi]-\{1\}$},\\
			\Omega &\text{ if $x=1$ }.
		\end{cases}
	\]
	Therefore the point $(1,\Omega)$ is in the graph  $\Gamma(f)$, contradicting the computability of $\Gamma(f)$.
	(see \cite{b:Braverman-Complexity-Real-Functions} for more details), i.e. $f$ is not graph-computable. On the other hand, $f\in\lp$
	because $\{s_n\equiv 0 \}_{n\geq 0}$ is a family of computable rational step functions approximating $f$.

	For the other direction take any two disjoint measure non-zero dense sets $A,B\subset [0,2\pi]$ such that $A\cup B = [0,2\pi]$. Define
	\[
		g(x)=
		\begin{cases}
			0 &\text{ if $x\in A$},\\
			1 &\text{ if $x\in B$ }.
		\end{cases}
	\]
	$g$ is easily seen to be graph computable because its graph is. On the other hand, because $A,B$ are both dense and of measure non zero,
	it is impossible to approximate $f$ with  rational step functions. \qed
	\end{proof}
}%end comment

\section{Baire Categories on $\lp$}
	Baire categories yield a topological size notion for function spaces. Intuitively a class of functions is meager if it is full of ``holes".
	Let us give a precise definition.

	\begin{definition}
	Given any rational step function $\psi$ and any rational $\epsilon >0$, we define the open ball centered in $\psi$ and of radius $\epsilon$ by	
	$$B(\psi,\epsilon)=\{f:[0,2\pi]\rightarrow \R| \ \nop{f-\psi}< \epsilon \} \ .$$ We call such a set a rational ball. We denote by
	$\mca{B}$ the set of all rational balls.
	\end{definition}

	A constructor (also called a strategy) is a function $\alpha : \mca{B}\rightarrow\mca{B}$ such that
	for any $B\in \mca{B}$, $\alpha (B) \subseteq B$.
	A computable constructor is a constructor $\alpha : \mca{B}\rightarrow\mca{B}$ such that there is a TM which on 
	input any $B\in\mca{B}$ outputs $\alpha(B)$. We shall also consider indexed computable constructors, i.e. of the form 
	$\alpha : \mca{B} \times \N\rightarrow\mca{B}$ such that there is a TM $M$ such that $M(B,i)=\alpha(B,i)$.

	Constructors are used to testify that a meager set is full of ``holes".
	\begin{definition}
		A set $X\subseteq \mca{P}(\R^{[0,2\pi]})$  is $\lp$-meager if $X=\cup_{i\geq 0} X_i$ and there exists a computable indexed constructor
		$\alpha$ such that for any $i\in\N$ and for any rational ball $B$ 
		\[
			\alpha(B,i)\cap X_i=\emptyset.
		\]
	\end{definition}
	The $\alpha$ in the above definition is said to avoid $X$ or to testify the meagerness of $X$.

	Meagerness is preserved for a special case of unions, called $\lp$-unions. Here is a definition.
	\begin{definition}
	$X=\cup_{i\geq 0} X_i \subseteq \mca{P}(\R^{[0,2\pi]})$ 
	is an $\lp$-union of $\lp$-meager sets if there exists a computable indexed constructor
	$\alpha:\mca{B}\times\N\times\N\rightarrow\mca{B}$ such that for any $i\in\N$, $\alpha(\cdot , \cdot,i)$ testifies $X_i$'s meagerness.
	\end{definition}

	The following result states that $L^p$-computable Baire categories indeed yield a size notion on $\lp$, i.e. the whole space is not meager,
	any point in the space is meager, and meagerness is preserved by suitable unions.

	\begin{theorem}
		\begin{enumerate}
			\item	If $f\in\lp$ then $\{ f\}$ is $\lp$-meager.
			\item	$\lp$ is not $\lp$-meager.
			\item	If $X=\cup_{i\geq 0} X_i$ is a $\lp$-union of $\lp$-meager sets, then $X$ is $\lp$-meager.
		\end{enumerate}
	\end{theorem}
	\begin{proof}	
		We prove the first item. Let $f\in\lp$ and $\{s_n\}_{n\geq 0}$ be a computable family of rational step functions
		approximating $f$, i.e. 
		\[
			\nop{f-s_n}\leq 2^{-n} \qquad \text{and} \qquad s_{n}=\{(c^n_i, d^n_i)|\, 1\leq i \leq t_{n}\}. 
		\]
		We define a computable strategy $\alpha$ witnessing the meagerness of $\{f\}$. 
		Let $B=(\psi,\epsilon)$ be a rational ball with endpoints
		$\{(a_i,b_i)|\, 1\leq i \leq t_B\}$. Let $n$ be big enough such that 
		$2^{-n}<\frac{\epsilon}{8}$. 
		On input $B$, $\alpha$ outputs a rational ball
		$B'=(\psi',\epsilon')$ where
		\[
			\psi'=\{(a'_i,b'_i)|\, 1\leq i \leq t_{B'}\}
		\]
		$\epsilon'=\frac{2^{-n}}{8\pi}$, and
		$$\{a'_i\}_{1\leq i \leq t_{B'}}= \{a_i\}_{1\leq i \leq t_B}\cup \{c^n_i\}_{1\leq i <t_{n}}$$
		For $1\leq i \leq t_{B'}$, define $k(i),l(i)$ as the unique integers such that
		\[
			[a'_i,a'_{i+1}] \subseteq [a_{k(i)},a_{k(i)+1}] \qquad \text{and} \qquad
			[a'_i,a'_{i+1}] \subseteq [c_{l(i)},c_{l(i)+1}] \qquad\text{and let}\qquad
		\]
		
		\[	
			e_i = |b_{k(i)}-d_{l(i)}| \qquad \text{and} \qquad f_i = a'_{i+1}-a'_i \, .
		\]
		Note that for any $1\leq i <t_{B'}$
		\[
			\psi|_{[a'_i,a'_{i+1}]}=b_{k(i)} \qquad \text{and} \qquad
			s_{n}|_{[a'_i,a'_{i+1}]}=d_{l(i)}
		\]
		For $1\leq i <t_{B'}$ let 
		\[
			b'_i=
			\begin{cases}
				b_{k(i)}   &\text{ if $e_i>4\cdot \frac{2^{-n}}{2\pi}$},\\
				d_{l(i)}+2\cdot 2^{-n}	&\text{ if $e_i<4\cdot \frac{2^{-n}}{2\pi}$ and $b_{k(i)}\geq d_{l(i)}$}\\
				d_{l(i)}-2\cdot 2^{-n}	&\text{ if $e_i<4\cdot \frac{2^{-n}}{2\pi}$ and $b_{k(i)}< d_{l(i)}$}		
			\end{cases}
		\]
		Let us see that $B'\subset B$. Let $g\in B'$ i.e. $\nop{g-\psi'}< \epsilon'$.
		Let $1\leq i \leq t_{B'}$ and suppose $e_i>4\cdot 2^{-n}$. 

		Case: $e_i> 4 \cdot 2^{-n}$
		\begin{align*}
			\int_{[a'_i, a'_{i+1}]} |g - \psi| \, dx 
			&\leq  \int_{[a'_i,a'_{i+1}]} |g - \psi'| \, dx +\int_{[a'_i,a'_{i+1}]} |\psi' - \psi| \, dx \\
			&\leq \epsilon' \cdot f_i + 0 \, .
		\end{align*}
		Case: $e_i> 4 \cdot 2^{-n}$
		\begin{align*}
			\int_{[a'_i, a'_{i+1}]} |g - \psi| \, dx 
			&\leq  \int_{[a'_i,a'_{i+1}]} |g - \psi'| \, dx +\int_{[a'_i,a'_{i+1}]} |\psi' - \psi| \, dx \\
			&\leq f_i (\epsilon' \cdot  +\frac{2^{-n}}{\pi}) \, .
		\end{align*}
		Thus 
		\begin{align*}
			\nop{g-\psi} 
			&= \sum_{i=1}^{t_{B'}-1} \int_{[a'_i, a'_{i+1}]} |g - \psi| \, dx \\
			&\leq \sum_{i=1}^{t_{B'}-1} f_i (\epsilon' + \frac{ 2^{-n}}{\pi} )\\ 
			&= (\frac{2^{-n}}{8\pi} + \frac{ 2^{-n}}{\pi})\sum_{i=1}^{t_{B'}-1} f_i\\
			&= \frac{2^{-n}}{4} + 2\cdot 2^{-n}\\
			&< \epsilon
		\end{align*}
		i.e. $g\in B$.

		Let us show that $B'\cap \{f\} = \emptyset$. Let $g\in B'$ i.e. $\nop{g-\psi'}< \epsilon'$.
		Let $1\leq i \leq t_{B'}$. We have
		\begin{align*}
			\int_{[a'_i, a'_{i+1}]} |g - s_{n}| \, dx 
			&\geq  \int_{[a'_i,a'_{i+1}]} |\psi'-s_{n}| \, dx -\int_{[a'_i,a'_{i+1}]} |g - \psi'| \, dx \\
			&> (\frac{2^{-n}}{\pi}-\epsilon') \cdot f_i \, .
		\end{align*}
		Thus 
		\begin{align*}
			\nop{g-s_{n}} 
			&= \sum_{i=1}^{t_{B'}-1} \int_{[a'_i, a'_{i+1}]} |g - s_{n}| \, dx \\
			&> \sum_{i=1}^{t_{B'}-1} f_i (\frac{2^{-n}}{\pi}-\epsilon')\\ 
			&= (\frac{2^{-n}}{\pi}-\frac{2^{-n}}{8\pi})\sum_{i=1}^{t_{B'}-1} f_i\\
			&> 2^{-n}
		\end{align*}
		i.e. $g\not\in \{f\}$, which ends the proof of the first statement of the theorem. The second statement will be proved 
		in Corollary \ref{c.toutpasmaigre}. The third one is left to the reader.
		\qed
	\end{proof}		
	
\subsection{The Banach-Mazur Game Characterization}

	In the classical theory of Baire categories (see \cite{b:oxtoby}), and also in the BC-model 
	\cite{b:lutz-baire-category-for-real-functions}, there is an alternative characterization of meagerness
	by Banach-Mazur games.
	Informally speaking, a Banach-Mazur game is a game between two strategies $\alpha$ and $\beta$,
	where the game begins with  some rational ball $B$.
	Then  $\beta \circ \alpha$ is applied successively on $B$.
	Such a game yields a unique function
	called the result of the game between $\alpha$ and $\beta$.
	A strategy $\beta$ wins the game against a class of functions $X$ if it can force the result of the game
	starting with any $\alpha$ and $B$ to be a function not in $X$.
	It is a classical result that the existence of a winning strategy against $X$ is equivalent to the meagerness of $X$.
	In the following section, we show that this alternative characterization also holds for $L^p$-computable Baire categories.

	Given two indexed constructors $\alpha,\beta$ where $\beta$ is a shrinking strategy (i.e. for every rational ball $B\in \mca{B}$,
	the radius of $\beta(B)$ is less than half the radius of $B$),
	the Banach-Mazur game between $\alpha,\beta$ proceeds in infinitely many rounds, starting with a rational ball 
	$B$, where $\alpha$ and $\beta$ are applied successively; i.e. $R_0 = \beta(\alpha(B,0),0)$,
	and round $i\in \Z^+$ corresponds to the rational ball $R_i=\beta(\alpha(R_{i-1},i),i)$. 
	The result of the game between $\alpha,\beta$ with initial ball $B$, denoted $R(\alpha,\beta,B)$ is the unique $\lpm$ function
	$f$, such that $f\in R_i$ for all $i\in\N$.

	Given a shrinking indexed constructor $\beta$, and a set $X\subseteq \R^{[0,2\pi]}$,
	$\beta$ is said to win the Banach-Mazur game against $X$, if for any indexed constructor $\alpha$, and any rational ball $B\in\mca{B}$,
	$R(\alpha,\beta,B) \not\in X$.

	The following result states that if both strategies are $L^p$-computable, then the resulting function also is.

	\begin{theorem}\label{t.const-comp-implies-comp}
		Let $\alpha,\beta$ be two computable indexed constructors, with $\beta$ shrinking, and let $B\in\mca{B}$.
		Then $R(\alpha,\beta,B)\in \lp$.
	\end{theorem}
	\begin{proof}
		Let $\alpha,\beta,B$ be as above, and denote by $r$ the radius of $B$. 
		Let 
		\[
			R_0= \beta(\alpha(B,0),0) \qquad\text{and}\qquad  R_i= \beta(\alpha(R_{i-1},i),i) \qquad (i\in\Z^+).
		\]
		For $n\in\N$, let 
		\[
			(s_{n},\epsilon_{n}) = R_{n+r}.
		\] 
		Thus $\{s_{n}\}_{n\geq 0}$ is a family of uniformly computable
		rational step functions,
		and $\epsilon_{n} < 2^{-n}$, because $\beta$ is shrinking. 
		Let $f=R(\alpha,\beta,B)$, i.e. $\forall i\in\N: f\in R_i$, thus for every $n\in\N$
		\[
			\nop{f-s_{n}}<2^{-n}
		\]
		i.e. $f\in\lp$.
		\qed
	\end{proof}

	The following result states that the classical characterization of meagerness by Banach-Mazur games also holds
	for $L^p$-computable Baire categories.
	
	\begin{theorem}
		Let $X\subseteq\R^{[0,2\pi]}$. 
		$X$ is $\lp$-meager iff there exists a shrinking $\lp$-computable constructor $\beta$ that wins the Banach-Mazur game against $X$.
	\end{theorem}
	\begin{proof}
		``$\Rightarrow$''.   Suppose $X$  is $\lp$-meager,
		i.e. $X=\cup_{i\geq 0}X_i$ and there exists an indexed computable constructor $\gamma$, such that for any $B\in\mca{B}$ and for every $i\in\N$,
		\[
			\gamma(B,i)\cap X_i=\emptyset .
		\]
		Let $A\in B$, $i\in\N$ and denote by $r$ the radius of $A$. Let us define $\beta(A,i)$
		\[
			\gamma(A,i)=B(\psi,\epsilon)
		\] 
		be a rational ball with
		center $\psi$ and radius $\epsilon$.
		Let 
		\[
			\epsilon'=\min\left\{ \frac{\epsilon}{2},\frac{r}{2}\right\} .
		\]
		Define
		\[
			\beta(A,i)=B(\psi,\epsilon').
		\]
		Let us show $\beta$ wins the Banach-Mazur game against $X$. 
		Let $\alpha$ be any indexed constructor and let $C\in\mca{B}$. Let 
		\[
			f=R(\alpha,\beta,C)
		\]
		i.e. for any $i\in\N$ 
		\[
			f\in\beta(\alpha(R_{i-1},i),i)
		\] 
		where $R_{i-1}$ is the ball obtained at round $i-1$.
		Letting $O=\alpha(R_{i-1},i)\in\mca{B}$, we have
		\[
			\beta(O,i) \subseteq \gamma(O,i) \qquad\text{and}\qquad \gamma(O,i)\cap X_i = \emptyset \, .
		\]
		Thus $f\not\in X_i$ for every $i\in\N$, i.e. $f\not\in X$.

		``$\Leftarrow$''. Let $\beta$ be as above.	We construct $\{X_i\}_{i\geq 1}$
		and a computable indexed constructor $\gamma$ such that $X=\cup_{i\geq1}X_i$ and $\gamma(\cdot,i)$ witnesses $X_i$'s meagerness.
		We have
		\begin{equation}\label{e.aff}	
			(\forall f \in X)  (\exists O \in\mca{B}:  f \in O)  (\exists j\in\N)  (\forall O'\subseteq O): f \not\in \beta(O',j) \, .
		\end{equation} 
		Otherwise it would be the case that 
		\begin{equation}\label{e.caff}
			(\exists f \in X) (\forall O \in\mca{B}: f\in O) (\forall j\in\N )( \exists O'\subseteq O): f \in \beta(O',j)
		\end{equation}
		Consider the following strategy $\alpha$. Let $f\in X$ be given by (\ref{e.caff}).
		For any $O\in\mca{B}$, $j\in\N$ let
		$O'$ be given by (\ref{e.caff}). Let $\alpha(O,j)=O'$. Let $S\in\mca{B}$ such that $f\in S$.
		By (\ref{e.caff}), $R(\alpha,\beta,S)=f$ which contradicts the assumption on $\beta$, i.e. (\ref{e.aff}) holds.

		Let 
		\[
			j\mapsto (O_{b(j)},b'(j))	
		\]
		Be a computable bijection between $\N$ and $\mca{B}\times\N$.
		For $i\in\N$ let $X_i$ be the set of $f\in X$ such that (\ref{e.aff}) holds with
		$O=O_{b(i)}$ and $j=b'(i)$. Clearly $X=\cup_{i\geq 1}X_i$.
		
		Let $O\in\mca{B}, i\in\N$. Denoting $O_{b(i)}$ by $O_{i}$, (\ref{e.aff}) yields
		\begin{equation}\label{e.aff2}	
			\forall f \in X_i: f\in O_i \text{ and } \forall O'\subseteq O_i: f \not\in \beta(O',b'(i)) \, .
		\end{equation} 
		We construct $\gamma(O,i)$ witnessing $X_i$'s meagerness. Suppose 
		\begin{equation}\label{e.aff3}
			O=(\{(a_j, b_j)|\, 1\leq j\leq t_O\},\epsilon) \qquad\text{and}\qquad O_i=(\{(c_j, d_j)|\, 1\leq j\leq t_i\},\epsilon_i)
		\end{equation}
		and denote by $\psi_O$ and $\psi_i$ the corresponding rational step functions.
		Compute $d=\nop{\psi_O-\psi_i}$.

		Case: $d>\epsilon_i$, i.e. $d=\epsilon_i+\nu$, where $\nu>0$.
		Let $S=B(\psi_O,r)\in\mca{B}$ where 
		\[
			r=\min \{\epsilon, \nu\} .
		\]
		Clearly $S\subseteq O$. Moreover $S\cap O_i=\emptyset$ because if $g\in S\cap O_i$,
		we have 
		\[
			\nop{\psi_O-\psi_i}\leq\nop{\psi_O-g}+\nop{g-\psi_i}<\epsilon_i + r \leq \epsilon_i +\nu =d
		\]
		a contradiction. Let $\gamma(O,i)=S$; by \ref{e.aff3} $\gamma(O,i)$ avoids $X_i$.

		Case $d<\epsilon_i$, i.e. $d=\epsilon_i-\nu$, where $\nu>0$.
		Let $S=B(\psi_O,r)\in\mca{B}$ where 
		\[
			r=\min \left\{\frac{\epsilon}{4}, \nu\right\}
		\]
		Clearly $S\subseteq O$. Moreover $S\subseteq O_i$ because if $g\in S$,
		we have 
		\[
			\nop{g-\psi_i}\leq\nop{g-\psi_O}+\nop{\psi_O-\psi_i}<\nu + d = \epsilon_i 
		\]
		i.e. $g\in O_i$.
		Let $\gamma(O,i)=\beta(S,b'(i))$. $\gamma(O,i)\subseteq O$ because $S\subseteq O$ and $\beta$ is a constructor.
		By (\ref{e.aff2}), $X_i\cap \gamma(O,i)=\emptyset$.

		Case: $d=\epsilon_i$. Let $S=(\{e_j,f_j| \, 1\leq j\leq t_S\} ,\epsilon')$, where
		\[
			\{e_j\}_{1\leq j \leq t_S} = \{a_j\}_{1\leq j \leq t_O} \cup \{c_j\}_{1\leq j \leq t_i}.
		\]
		Let $1\leq j\leq t_S$. Let $l(j),k(j)$ be the unique integers such that
		\[
			I_j := [e_j,e_{j+1}] \subseteq [a_{l(j)},a_{l(j)+1}] \qquad\text{and}\qquad 
			[e_j,e_{j+1}] \subseteq [c_{k(j)},c_{k(j)+1}]
		\]
		i.e.
		\[
			\psi_O|_{I_j} = b_{l(j)} \qquad\text{and}\qquad \psi_i|_{I_j} = d_{k(j)} \, .
		\]
		Let 
		\[
			f_j=
			\begin{cases}
				b_{l(j)} - \frac{\epsilon}{4\pi} &\text{if $b_{l(j)}\leq d_{k(j)}$}\\
				b_{l(j)} + \frac{\epsilon}{4\pi} &\text{if $b_{l(j)}> d_{k(j)}$}
			\end{cases}
		\]
		and denote by $\psi_S$ the step function associated to $S$.
		We have 
		\begin{align*}
			\nop{\psi_S-\psi_i} &= \sum_{j=1}^{t_S}\int_{I_j} |\psi_S(x)-\psi_i(x)|\, dx \\
			&= \sum_{j=1}^{t_S}\left[ \int_{I_j} |b_{l(j)}-d_{k(j)}|\, dx + \int_{I_j} \frac{\epsilon}{4\pi}\, dx \right]\\
			&= \sum_{j=1}^{t_S}\left[ \int_{I_j} |\psi_O(x)-\psi_i(x)|\, dx + \frac{\epsilon}{4\pi}\cdot |I_j|\right] \\
			&= \nop{\psi_O-\psi_i} +\frac{\epsilon}{2}\\
			&+ \epsilon_i + \frac{\epsilon}{2} \, .
		\end{align*}
		Similarly 
		\[
			\nop{\psi_S-\psi_O} = \frac{\epsilon}{2} \, .
		\]
		$S\cap O_i=\emptyset$, because if $g\in S$, then
		\[
			\nop{g-\psi_i} \geq \nop{\psi_S-\psi_i} - \nop{\psi_S-g} = \epsilon_i +\frac{\epsilon}{2} -\frac{\epsilon}{4} > \epsilon_i
		\]
		i.e. $g\not\in O_i$.
		Moreover $S\subseteq O$, because if $g\in S$, then
		\[
			\nop{g-\psi_O} \leq \nop{g-\psi_S} +\nop{\psi_S-\psi_O} < \frac{\epsilon}{4} + \frac{\epsilon}{2} < \epsilon
		\]
		i.e. $g\in O$.
		Therefore putting $\gamma(O,i)=S$ avoids $X_i$ similarly to the first case.
		\qed
	\end{proof}
	
	This alternative characterization makes it easy to prove that the whole space is not meager,
	as the following result shows.

	\begin{corollary}\label{c.toutpasmaigre}
		$\lp$ is not $\lp$-meager.
	\end{corollary}
	\begin{proof}
		Suppose $\lp$ is $\lp$-meager and let $\beta$ be a shrinking indexed strategy winning the BM game against $X$.
		Let $\alpha$ be a computable indexed constructor, and let $B\in\mca{B}$.
		Then $R(\alpha,\beta,B)\in \lp$ by Theorem \ref{t.const-comp-implies-comp}, a contradiction.
		\qed
	\end{proof}

\section{Convergence of Fourier Series for Functions in $\lp$}

	In this section we investigate how well can  Fourier series approximate $L^p$-computable functions. Fourier series are trigonometric 
	series that are broadly used for the approximation of arbitrary periodic functions in many different areas including electrical engineering,
	signal and image processing, and data compression. 

	It is a classical result \cite{b:zygmund.trigo.series.vol.1,b:hardy.Fourier.series} that  for any $p>1$,  the Fourier series of any function
	in $L^p$-converges to $f$ in the $L^p$ norm, therefore this also holds for $L^p$-computable functions.
	
	For $L^1$, the situation is different:  
	Kolmogorov \cite{b:kolmo.counter-example.in.L1} constructed
	a function $f\in L^1$ whose Fourier series diverges almost everywhere. Using some of his techniques we show a stronger,
	\emph{typical} result for the computable case, namely that the class of $L^1$-computable
	functions captures some very ``complicated'' functions, so that most of them cannot be approximated by their Fourier series;
	more precisely we show that 
	the set of $L^1$-computable functions whose Fourier series does not diverge a.e. is $\lun$-meager, i.e. negligible.
	As  corollaries, our result implies the existence of a  function in $\lun$ whose Fourier series diverges almost everywhere,
	as well as Kolmogorov's result \cite{b:kolmo.counter-example.in.L1}. 

	First let us give some notation.
	The Fourier series of any function $f\in \R^{[0,2\pi]}$ is given by 
	\[
		S[f]= \frac{a_0}{2} +\sum_{n=1}^{\infty} (a_n \cos nx + b_n \sin nx)
	\]
	where 
	\[
		a_n = \frac{1}{\pi}\int_0^{2\pi} f(t)\cos nt \, dt, \quad b_n = \frac{1}{\pi}\int_0^{2\pi} f(t)\sin nt \, dt
		\quad (n=1,2,\ldots).
	\]
	$S_l(f,x)$ denotes the partial $l$th sum of the Fourier series of $f$ evaluated at $x$.

	The following is the main result of this section.

\begin{theorem}\label{t.computable.kolmogorov}
	The set $\{ f\in \R^{[0,2\pi]} |$  $S[f]$ does not diverge a.e.$\}$
	is $\lun$-meager.
\end{theorem}
\begin{proof}
	We need the following Lemma. Kolmogorov  \cite{b:kolmo.counter-example.in.L1}  proved a $L^1$ version of it. 
	A slight modification of his proof makes it hold in $\lun$.
	\begin{lemma}\label{l.polytrigo} 
		There exists a family $\{ f_{n} \}_{n\geq 0} \in \lun$ of positive trigonometric polynomials with constant term
		$\frac{1}{2}$ together with a family of sets $\{ E_{n} \}_{n\geq 0}$ and a constant $A$	such that:
		\begin{enumerate}
			\item	$\lim_{n\rightarrow\infty}|E_{n}|= 2\pi$
			\item	For $x\in E_{n}$ there exists $1\leq j\leq n$ such that:
					$|S_{m_j}(f_{n},x)|> \log^{\frac{1}{2}} n -A$
		\end{enumerate}
	\end{lemma}
	\begin{proof}
	The proof of Lemma \ref{l.polytrigo} we shall give is based on \cite{b:zygmund.trigo.series.vol.1}.
	Let $n\in \N$ and let $a_j =\frac{4\pi j}{2n+1} \quad(j=1,2,\ldots , n)$. Let $\Delta_i'= (a_i - n^{-2}, a_i + n^{-2} )$.
	Let $$f_{n}(x)= \frac{1}{n} \sum_{i=1}^nK_{m_i}(x-a_i)$$ 
	where $m_1=n^4$, and the $m_j$'s ($1< j\leq n$) are defined recursively as the smallest integer satisfying
	$$m_{j+1}>2m_j, \text{ and } 2m_j +1 \equiv 0 \pmod{2n+1} $$ 
	where $K_l(t)$ are the Fejer kernels given by
	$$K_l(t)= \frac{2}{l+1} \left[\frac{\sin\frac{1}{2}(l+1)t}{2 \sin \frac{1}{2}t}\right]^2 .$$

	Clearly $\{f_{n}\}_{n} \in \lun$, $f_{n} \geq 0$ and the constant term of $f_{n}$ is $\frac{1}{2}$.
	We have
	$$S_{m_j}(f_{n},x)= \frac{1}{n}\sum_{i=1}^j K_{m_i}(x-a_i) + \frac{1}{n}\sum^n_{i=j+1}\left[ \frac{1}{2}+ \sum_{l=1}^{m_j}
	\frac{m_i-l+1}{m_i+1}\cos l(x-a_i) \right] ,$$
	and since 
	$$m_i-l+1 = (m_i-m_j) +(m_j-l+1),$$
	we have
	$$S_{m_j}(f_{n},x)= \frac{1}{n}\sum_{i=1}^j K_{m_i}(x-a_i) + 
	\frac{1}{n}\sum^n_{i=j+1} \frac{m_j+1}{m_i+1}K_{m_j}(x-a_i)
	+\frac{1}{n}\sum^n_{i=j+1} \frac{m_i-m_j}{m_i+1}D_{m_j}(x-a_i) ,
	$$
	where $D_l$ are the partial sums given by 
	$$D_l(x)=\frac{1}{2}+\sum_{\nu = 1}^l\cos\nu x = \frac{\sin (l+\frac{1}{2})x}{2\sin\frac{1}{2}x} .$$

	Let $\Delta_i=(a_{i-1},a_i), \ \Delta_i'=(a_{i}-n^{-2},a_i+n^{-2}) \ (i=1,\ldots, n) $.
	The estimate $K_m(t)=O(m^{-1}t^{-2})$, together with $m_i\geq n^4$, shows that $K_{m_i}(x-a_i)$ is uniformly
	bounded outside $\Delta'_i$, and so the contribution of the first two terms on the right in the last formula for
	$S_{m_j}(f_{n},x)$ is less than an absolute constant $A$ outside of $\cup\Delta_i'$:
	\begin{equation}\label{e.3.8}
		S_{m_j}(f_{n},x)\geq \frac{1}{n} \left| \sum_{i=j+1}^{n} \frac{m_i-m_j}{m_i+1}D_{m_j}(x-a_i) \right|-A \qquad (j=1,2,\ldots,n;\ x\not\in\cup\Delta_i').
	\end{equation}
	Since
	$$(2m_j+1)\frac{1}{2}a_i=\frac{2m_j+1}{2n+1}2i\pi \equiv 0 \pmod{2\pi},$$
	we have
	$$D_{m_j}(x-a_i) = \frac{\sin(m_j+\frac{1}{2})(x-a_i)}{2\sin \frac{1}{2}(x-a_i)}= \frac{\sin(m_j+\frac{1}{2})x}{2\sin \frac{1}{2}(x-a_i)} .$$
	
	Suppose that $x\in\Delta_j$, $1\leq j \leq n-\sqrt{n}$. Due to $m_{i+1}\geq 2m_i+1$, the multipliers of the $D_{m_j}$ in (\ref{e.3.8})
	are not less than $\frac{1}{2}$, and since the denominators $2\sin\frac{1}{2}(x-a_i)$ are all of constant sign for $i>j$ we get
	\begin{align*}
		\frac{1}{n}\left| \sum_{i=j+1}^{n} \frac{m_i-m_j}{m_i+1}D_{m_j}(x-a_i) \right|
		&\geq \frac{1}{2n} |\sin (m_j+\frac{1}{2})x| \sum_{i=j+1}^{n} \frac{1}{a_i-a_{j-1}}\\
		&= \frac{1}{2n} |\sin (m_j+\frac{1}{2})x| \frac{2n+1}{4\pi}\sum_{k=2}^{n-j+2} \frac{1}{k}\\
		&\geq \frac{1}{4\pi} |\sin (m_j+\frac{1}{2})x| (-1+\log (n-j))\\
		&\geq \frac{1}{4\pi} |\sin (m_j+\frac{1}{2})x| (-1+ \frac{1}{2}\log n)\\
		&\geq \frac{1}{9\pi} |\sin (m_j+\frac{1}{2})x| \log n ,
	\end{align*}
	for $n$ large enough; thus at the points not in $\cup\Delta_i'$ for which
	\begin{equation}\label{e.3.10}
		\left|\sin (m_j+\frac{1}{2})x\right| \geq \frac{9\pi}{(\log n)^{\frac{1}{2}}},
	\end{equation}
	we have 
	\begin{equation}\label{e.3.11}
		|S_{m_j}(f_{n},x)|\geq (\log n)^{\frac{1}{2}} -A .
	\end{equation}
	
	The set of $x$ in $(0,2\pi)$ where  (\ref{e.3.10}) fails has measure $O(\log^{-\frac{1}{2}} n)$.
	Therefore, if from $(0,a_{[n-n^{\frac{1}{2}}]})$ we remove the points where (\ref{e.3.10}) fails, and those 
	which are in $\cup\Delta'_i$, and denote the remainder by $E_{n}$, then
	$$2\pi -E_{n} = O(\log^{-\frac{1}{2}}n) + O(n\cdot n^{-2})+ O(\sqrt{n}\cdot n^{-1}) = o(1).$$
	For each $x\in E_{n}$ and a suitable $j=j(x)$   (\ref{e.3.11}) holds, thus ending the proof of the Lemma.
	\qed
	\end{proof}
	
	Let us prove Theorem \ref{t.computable.kolmogorov} 
	Let $A$ be given by Lemma \ref{l.polytrigo} and let $A_{n} = \log^{\frac{1}{2}} n -A$.
	Let $\{n_k\}_k$ be a sequence of integers increasing rapidly enough to guarantee that
	$$\sum_k A_{n_k}^{-\frac{1}{2}} < \infty$$ for example $n_k=2^{2^k}$.
	$f_{n_k}-\frac{1}{2}$ has constant term zero. Hence define  $\{q_k\}_k$ to increase fast enough such that:
	the order of $f_{n_k}(q_kx)-\frac{1}{2}$ is strictly less than the order of $f_{n_{k+1}}(q_{k+1}x)-\frac{1}{2}$, i.e.
	they  do not overlap.

	Let $h$ be a strategy. Let us construct a strategy $g$, winning the BM game. Suppose the $m$'th round ($m$ uneven) of the Banach Mazur
	game between $h$ and $g$ yields the open set $O_m = (\psi_m, \epsilon_m )$, where $\psi_m$ is the  step function, given by
	$$\psi_m |_{[a_i^m,a^m_{i+1})} = b^m_i \qquad  (i=1,\ldots,t_m).$$
	By definition 
	$$O_m =\{ f :[0,2\pi]\rightarrow \R | \ \noun{ f-\psi_m } < \epsilon_m \} .$$

	Consider the function 
	$$e_{m+1}|_{[a_i^m,a^m_{i+1})} = F_{\alpha(m+1)}(x)-b^m_i \qquad (i=1,\ldots,t_m)$$ 
	where 
	$$F_k (x)= A_{n_k}^{-\frac{1}{2}}(f_{n_k}(q_k x)-\frac{1}{2})$$
	and $\alpha(m+1)$ will be determined later. Clearly $e_{m+1}\in \lun$ given $O_m$, therefore there exists a computable step
	function $\psi_{m+1}$ such that 
	$$\noun{ \psi_{m+1}-e_{m+1}} < \frac{\epsilon_{m}}{4}.$$ 
	Define $g(O_m,m+1)= (\psi_{m+1}, \epsilon_{m+1})$
	where 
	$$\epsilon_{m+1}= \min \left\{ \frac{\epsilon_m}{4},\beta(m+1) \right\}$$ with $\beta(m+1)$  to be determined later.

	Let us check that $O_{m+1} \subset O_m$. Let $v\in O_{m+1}$, i.e. $\noun{v-\psi_{m+1}} < \frac{\epsilon_m}{4}$.
	We have 
	\begin{align*}
		\noun{v-\psi_{m}} &\leq \noun{v-\psi_{m+1}} + \noun{\psi_{m+1}-\psi_m}\\
		&\leq \frac{\epsilon_m}{4} +\noun{\psi_{m+1}-e_{m+1}} + \noun{e_{m+1}- \psi_m}\\
		&\leq \frac{\epsilon_m}{2} +  \int_0^{2\pi} |e_{m+1}(x)-\psi_m(x)|\, dx \\
		&=  \frac{\epsilon_m}{2} +  \int_0^{2\pi} F_{\alpha (m+1)}(x)\, dx \\
		&=  \frac{\epsilon_m}{2} +  A^{-\frac{1}{2}}_{n_{\alpha(m+1)}}\int_0^{2\pi} [f_{n_{\alpha (m+1)}}(x)-\frac{1}{2}]\, dx \\
		&= \frac{\epsilon_m}{2} + 2\pi A^{-\frac{1}{2}}_{n_{\alpha(m+1)}} < \epsilon_m .
	\end{align*}
	Choosing $\alpha(m+1)$ big enough (depending on $\epsilon_m$) ensures the last inequality. 
	Let 
	\begin{align*}
		G_{m+1} &= \{ x | \ q_{\alpha(m+1)} x \in E_{n_{\alpha(m+1)}} \}\\
		&= \{ x | \ \exists m_j \ (j=1,\ldots,n_{\alpha(m+1)}) \text{ such that } 
		S_{m_j}(f_{n_{\alpha(m+1)}},q_{\alpha(m+1)}x) > A_{n_{\alpha(m+1)}}  \}\\
	\end{align*}
	We have 
	$$\lim_{m\rightarrow\infty}|G_{m+1}|=\lim_{m\rightarrow\infty}|E_{n_{\alpha(m+1)}}|=2\pi .$$
	If $x\in G_{m+1}$ then there exists $1\leq j \leq n_{\alpha(m+1)}$ such that $S_{m_j}(F_{\alpha(m+1)},x)> A^{\frac{1}{2}}_{n_{\alpha(m+1)}}$,
	i.e. 
	$$S_{m_j}(\psi_{m+1},x)> A^{\frac{1}{2}}_{n_{\alpha(m+1)}}-b_i^m > m+1$$
	where choosing $\alpha(m+1)$ big enough (depending on $b_i^m \ (1\leq i \leq t_m)$) ensures the last inequality.

	Let $\psi=R(h,g,O_0)$ be the result of the BM game between $h$ and $g$, in particular 
	$$\noun{\psi - \psi_{m+1}}<\beta(m+1)  \qquad (m  \text{ uneven}).$$

	\begin{lemma} \label{l.inegal.fourier}
		For any two functions $p,q\in \mathrm{L}^1$ and any $\epsilon > 0$, $l\in\N$, such that $\noun{p-q}<\epsilon$, we have
		$$|S_l(p,x)-S_l(g,x)|\leq l\epsilon$$
	\end{lemma}
	\begin{proof}
		Let $p,q,\epsilon,l$ be as above.
		\begin{align*}
			|a_{n}(p)-a_{n}(q)| &= \frac{1}{2\pi} \left| \int^{2\pi}_0 p(x) \cos nx \, dx - \int^{2\pi}_0 q(x) \cos nx \, dx \right|\\
			&\leq \frac{1}{2\pi}  \int^{2\pi}_0 |p(x)-q(x)|  \, dx 
			= \noun{p-q}
		\end{align*}	
	The same arguments holds for the $b_{n}$'s coefficients of the Fourier series.

	Therefore 
		\begin{align*}
			|S_l(p,x)-S_l(q,x)| &= \left|\sum_{n=1}^l [a_{n}(p) \cos nx + b_{n}(p)\sin nx] - \sum_{n=1}^l [a_{n}(q) \cos nx + b_{n}(q)\sin nx] \right|\\
			&\leq \sum_{n=1}^l |a_{n}(p)-a_{n}(q)|  + |b_{n}(p)-b_{n}(q)| \\
			&\leq l\epsilon
		\end{align*}	
	which proves the Lemma.
	\end{proof}

	Let us show that for almost every $x$, the Fourier series of $\psi$ diverges.
	For every uneven $m$ and for every $x\in G_{m+1}$ there exists $j$ such that
	$S_{m_j}(\psi_m,x)> m+1$.
	Moreover $\noun{\psi - \psi_{m+1}} < \beta(m+1)$, thus by Lemma \ref{l.inegal.fourier}
	$$S_{m_j}(\psi,x) \geq S_{m_j}(\psi_m,x)-m_j \beta(m+1) > \frac{m+1}{2}$$
	where the last inequality holds by an appropriate choice of $\beta(m+1)$ (depending on $m_j$ where $1\leq j \leq n_{\alpha(m+1)}$).
	The fact that $\lim_{m\rightarrow\infty}|G_{m+1}|=2\pi$ ends the proof.
	\qed

\end{proof}

\begin{corollary}
	There exists $f\in\lun$ whose Fourier series diverges almost everywhere.
\end{corollary}

\begin{corollary}[Kolomogorov \cite{b:kolmo.counter-example.in.L1}]
	There exists $f\in L^1$ whose Fourier series diverges almost everywhere.
\end{corollary}

\noindent
\textbf{Acknowledgments.} I thank E. Mayordomo and M. L\'opez-Vald\'es for helpful discussions.

%\end{spacing}{3} % double space for easy correction
%%%%%%%% bibliography%%%%%%%%%%%%%%%

\end{document}